\documentclass[11pt,english]{article}
\usepackage[T1]{fontenc}
\usepackage[latin9]{inputenc}
\usepackage{geometry}
\usepackage{color}
\definecolor{page_backgroundcolor}{rgb}{1, 1, 1}
\pagecolor{page_backgroundcolor}
\usepackage{babel}
\usepackage{amsthm}
\usepackage{amsmath}
\usepackage{graphicx}
\usepackage[numbers]{natbib}
\usepackage[unicode=true]
 {hyperref}

\makeatletter

\providecommand{\tabularnewline}{\\}

\numberwithin{equation}{section}
\numberwithin{figure}{section}

\usepackage{longtable}
\usepackage{fancyhdr}
\makeatother

\begin{document}

\title{ARL Libraries and Research: Correlates of Grant Funding}

\author{Ryan P. Womack%
\thanks{Ryan Womack is Data Librarian at Rutgers University-New Brunswick,
rwomack@rutgers.edu%
}}
\maketitle
\begin{abstract}
While providing the resources and tools that make advanced research
possible is a primary mission of academic libraries at large research
universities, many other elements also contribute to the success of
the research enterprise, such as institutional funding, staffing,
labs, and equipment. This study focuses on U.S. members of the Association
for Research Libraries (ARL). Research success is measured by the
total grant funding received by the University, creating an ordered
set of categories. Combining data from the NSF\textquoteright s National
Center for Science and Engineering Statistics, ARL Statistics, and
IPEDS, the primary explanatory factors for research success are examined.
Using linear regression, logistic regression, and the cumulative logit
model, the best-fitting models generated by ARL data, NSF data, and
the combined data set for both nominal and per capita funding are
compared. These models produce the most relevant explanatory variables
for research funding, which do not include library-related variables
in most cases. 

Keywords: Academic Libraries, Research, Universities\pagebreak{}
\end{abstract}

\section{Background and Literature Review}

Academic libraries are under increasing pressure to demonstrate their
relevance to the scholarly enterprise via concrete metrics. The literature
of professional librarianship is replete with discussions the importance
of libraries, but thorough quantitative studies are somewhat rarer.
Several quantitative approaches to evaluating the impact of academic
libraries have been used, as discussed in the literature review below. 

Some studies demonstrate the importance of the library to student
outcomes. Whitmire \citep{Whitmire2002} found that gains in critical
thinking skills among undergraduates, as measured by the College Student
Experiences Questionnaire, were linked to library measures taken from
the Integrated Postsecondary Education Data System (IPEDS). Mezick
\citep{Mezick2007} also used IPEDS data along with data from the
Association of Research Libraries (ARL) and the Association of College
and Research Libraries (ACRL) to show a correlation between library
expenditures and professional staff and student retention. Researchers
at the University of Minnesota \citep{SoriaFransenNackerud} used
detailed student records to demonstrate a positive relationship between
academic performance and library use.

A second approach has been to look for the impact of library resources
on faculty publications. For example, Budd \citep{Budd2006} studies
faculty productivity and uses rank-order correlations to show a moderate
association between the quantity of faculty publishing at ACRL institutions
and library expenditure and volumes held. The number of PhD's awarded
also shows similar levels of correlation. Surveys of faculty attitudes
towards academic libraries, such as Mikitish and Radford \citep{MikitishRadford},
are another way to establish value.

Hendrix \citep{Hendrix2010} used principal components analysis to
study the relationship between faculty citations and library variables
from the ARL Statistics. While strong assocations were present in
the initial dataset, no associations with faculty citations were found
when using size-independent measures of library activity. In an earlier
article, Hendrix \citep{Hendrix2008} also conducted a bibliometric
study on medical schools using principal components methodology.

Another way of addressing the pressure to demonstrate the continuing
relevance of libraries is to adopt business paradigms, such as return
on investment (ROI) \citep{Coyle2006}. In contrast to a business
environment with clearly defined profit and loss, the inputs and outputs
in the library context are harder to pin down, and are imperfectly
addressed by existing data sources. Tenopir \citep{Tenopir2010} described
the evalution of ROI by working with administrators to understand
their attitudes towards library support and its impact on grant funding.
At several institutions, the article citations used in grant proposals
were studied and combined with qualitative information from surveys
of faculty submitting grant proposals which testified to the value
of the library.

Turning to studies that use larger data sets and more extensive quantitative
methods, Allen and Dickie \citep{AllenDickie} built a regression
model that relates library expenditure as the response variable to
various institutional measures such as the size of programs, enrollments,
and faculty. 

Weiner \citep{Weiner2009} built a dataset that combined IPEDS, ARL,
and \emph{US News and World Report} peer assessment scores, along
with several other sources to determine factors influencing institutional
reputation. She then used stepwise linear regression to build explanatory
models. Library expenditure was influential in all models, and grants
and instructional expenditures were also influential.

\section{Goals of this study}

A major characteristic and limitation of most of the studies mentioned
above (with the notable exception of Weiner) is that they use only
library data to explain the outcome of interest. But in the context
of a university, a well-performing library may be correlated with
many other factors that more directly influence student success, since
the best libraries are typically at the best schools with the best
funding, best support services, best faculty, and so on. Allen and
Dickie's work shows how library funding can be predicted from these
factors. Working only with library variables to demonstrate the library's
relevance does not allow for alternative explanations and is a weak
form of proof.

However, Weiner's study did include other institutional reputational
factors in order to select a model that combined variables from different
spheres. The present study takes a similar approach to modeling both
library and other academic factors, but with a wider range of statistical
methods and a larger selection of variables. This will provide one
method of determining whether library characteristics are the primary
explanatory factors for the outcome, or whether they are only secondary
factors that have some explanatory power due to their correlation
with other primary factors.

Our primary response variable will be research productivity, as measured
by grant funding. Grant funding for research is a central characteristic
for the reputation and identity of major research universities. We
will look at a representative group of research universities and assess
whether library or other academic and institutional characteristics
are related to grant funding. A secondary dimension of interest is
the effect of fitting linear regression, logistic regression for binary
outcomes, and cumulative logit models for multi-category ordered outcomes.
Logistic and cumulative logit methods can help explain data that is
categorical in nature, rather than continuous, and may provide a better
fit than linear regression in many settings. By comparing different
fitted models, we will begin to understand the variables that are
most closely related to research funding. Most importantly, our model
selection process will select the best explanatory variables from
among all candidate variables. Whether or not the final fitted model
includes library-related variables will be a strong indicator of library
relevance to the university's research productivity.

\section{Methods}

\subsection{Data Collection}

The Association of Research Libraries (ARL) is the leading grouping
of large research libraries in North America. The ARL Statistics have
been collected annually since 1908 \citep{ARL2012}. In 2012, there
were 125 members, 17 of which were in Canada. The 108 members located
in the United States consist of 99 university research libraries and
9 institutional libraries (e.g., the New York Public Library, National
Library of Medicine, Library of Congress, and so on). This study uses
data only from the 99 US university libraries, who compete for research
funding under similar conditions. The Canadian research funding environment
is not directly comparable.

Although the ARL membership contains most of the largest universities
from a research funding standpoint, there are notable exceptions.
Institutions that receive large research grants such as Stanford,
the California Institute of Technology, Carnegie Mellon, and others
are not ARL members. Other institutions in the ARL are ranked below
the top 200 in research funding, such as Howard University (\#208
in 2012) or Kent State University (\#248), far below many non-ARL
members. However, the ARL has the longest-running and most complete
collection of library statistics, and this data sample has the most
potential for detailed comparisons over time. With the exceptions
noted, it remains a very representative grouping of the most active
research universities. Data from the year 2012 was used for comparability
with the most recently available data at the time of the study, collected
from the other sources described below. 

The National Center for Science and Engineering Statistics (NCSES)
of the National Science Foundation (NSF) is the most systematic collection
of data on research funding and inputs to research in the US. The
Higher Education Research and Development (HERD) survey, which is
the most systematic collection of data on research funding and inputs
to research in the United States \citep{HERD2012}. The HERD reports
annually on levels of research funding from all sources: federal,
state, local, nonprofit, business, internal institutional funding,
and other sources. For the purposes of this study, the total research
funding received in 2012 was the primary response variable of interest,
although federal funding is the largest share of funding and closely
tracks the total.

The NCSES Survey on Science and Engineering Facilities \citep{SSEF2013}
reports on the total amount of existing square footage of research
space, as well as newly constructed space in the last year, dedicated
to science and engineering research at universities in the US, in
laboratories, animal research facilities, computer labs, equipment
rooms, and other such facilities. The latest available data, at the
time of the study, from the fiscal year 2011 was used. Data is collected
every two years, so there is no direct equivalent to 2012. Since these
variables function as a likely input to future grant receipts, using
the earlier year is reasonable. Planned construction and repair and
renovation costs were excluded from the dataset since they are not
likely to be directly related to grant success.

Finally, in order to add other measures of staffing and salary expenses
in non-library categories, along with additional institutional characteristics,
data for 2012 was extracted from the Integrated Postsecondary Education
Data System (IPEDS) of the National Center for Education Statistics
\citep{IPEDS2012}. The IPEDS data reports the number of employees,
the total salary expenses, and the number of Full-Time Equivalent
(FTE) employees in several categories. All of these ways of measuring
employment are included in the dataset.

Since medical research is a large component of overall research dollars,
data from the Association of Academic Health Sciences Libraries was
also considered \citep{AAHSL2014}. However, the overall magnitude
of medical library expenditures and staffing is not large compared
to their general academic library counterparts. For example, at the
University of Michigan, collections spending is \$2 million in the
medical library versus \$24 million in the main library, and professional
FTE employment is 15 versus 212. The medical data also has many missing
values and introduces questions of comparability that would require
investigation of each institution's library and institutional configuration
of its medical research vis-à-vis the rest of the campus. The IPEDS
data contains indicator variables for medical degree-granting and
presence of a hospital, so these can serve as a proxy for any distinctive
medical effects. Based on these considerations, the AAHSL data was
not included in the present study.

The ARL, NCSES, and IPEDS data described above were merged into a
single dataset for the 99 US ARL institutions under study. At this
stage there were 75 possible predictor variables representing inputs
from library, research, infrastructure, and general staffing characteristics
of the institutions. Details of the data cleaning process are described
in the Appendix.

The data files used in this study, along with the R code used to conduct
the analysis, are available from openICPSR at \href{http://doi.org/10.3886/E45486V1}{http://doi.org/10.3886/E45486V1}.
The R code provides more detail on the steps used in the modeling
process described below. The abbreviated variable names used to report
results in the paper correspond to those in the R code. Tables 4,
5, and 6 provide more complete descriptive names of the variables.

\subsection{Modeling }

The data is modeled along three different dimensions. First, we consider
as explanatory variables the library-based effects alone, then we
use the other academic institutional data alone, and then evaluate
the model with both library and institutional data. These approaches
will be termed \emph{library, academic, }and \emph{combined}.\emph{ }

Second, we want to understand whether continuous or categorical response
variables produce more effective models. We will develop linear regression
models, logistic regression models, and cumulative logit models for
a four-category breakdown of research funding. These will be termed
\emph{linear, binary, }and \emph{clm} in what follows. Institutions
are grouped into four categories of research funding, based on the
NCSES reported dollar amounts of research funding received in FY 2012,
are listed below.

\medskip{}

\begin{center}
\begin{tabular}{|c|c|c|}
\hline 
 & Research funding (in millions of \$) & \# of observations\tabularnewline
\hline 
\hline 
1 & <200 & 25\tabularnewline
\hline 
2 & 200-400 & 31\tabularnewline
\hline 
3 & 400-700 & 22\tabularnewline
\hline 
4 & >700 & 21\tabularnewline
\hline 
\end{tabular}
\par\end{center}

\medskip{}

These cutoff points were chosen as natural grouping points for the
data that provide roughly equal numbers of observations per category.
The binary categorization into Low or High research activity is generated
by the dividing line of \$400 million in research funding.

As discussed above, the previous library literature provides little
guidance on the modeling choices, so we rely on standard statistical
principles to make decisions. One rule of thumb is to allow no more
than 10 observations per predictor in order to achieve effective power
\citep{Peduzzi1996}. Therefore, developing models with 10 or even
fewer explanatory variables is a primary goal. Grouping the data into
binary or a limited number of categories increases power and allows
us to build more parsimonious models. 

Finally, our third dimension will be the nominal, or as reported amount
of research funding, versus per capita research funding. We will measure
the research funding received per faculty member, and build models
on this basis to understand how \emph{per capita} measures differ
from the \emph{nominal} for each of the model variants\emph{.}

\subsection{Library Data Univariate Analysis}

For the variables taken from the ARL data, we analyze the correlation
(using Spearman's rho) between continuous (\texttt{RD}) and categorical
measures of nominal research funding (\texttt{RDCAT} for 4 category,
\texttt{RDBIN} for binary), and each of the predictor variables. These
results are reported in Tables 1, 2, and 3.

\begin{table}[h]

\protect\caption{Correlations (Spearman's rho) between RDCAT, RDBIN, and RD}

\begin{tabular}{|c|c|c|c|}
\hline 
 & RDCAT & RDBIN & RD\tabularnewline
\hline 
\hline 
RDCAT & 1 & 0.8889726811 & 0.9658004777\tabularnewline
\hline 
RDBIN & 0.8889726811 & 1 & 0.8585702401\tabularnewline
\hline 
RD & 0.9658004777 & 0.8585702401 & 1\tabularnewline
\hline 
RDFED & 0.9310306616 & 0.827193919 & 0.9664687693\tabularnewline
\hline 
RDNSF & 0.6993151302 & 0.5897322164 & 0.7144836116\tabularnewline
\hline 
RDNSFMATH & 0.6417228031 & 0.5626362243 & 0.6464274781\tabularnewline
\hline 
region & -0.0731276165 & -0.0787972392 & -0.0553181013\tabularnewline
\hline 
membyr & -0.5398608226 & -0.4333554504 & -0.5343527263\tabularnewline
\hline 
vols & 0.5646221133 & 0.5391022438 & 0.5825974026\tabularnewline
\hline 
illtot & 0.4188450169 & 0.3244596837 & 0.4248237477\tabularnewline
\hline 
ilbtot & 0.2902671333 & 0.2396009972 & 0.3144094001\tabularnewline
\hline 
grppres & 0.3659092851 & 0.3897117848 & 0.3583211966\tabularnewline
\hline 
presptcp & 0.3598719112 & 0.3865004228 & 0.361448242\tabularnewline
\hline 
reftrans & 0.1733902042 & 0.1508249344 & 0.1784285229\tabularnewline
\hline 
initcirc & 0.4558879977 & 0.4813412891 & 0.4799134199\tabularnewline
\hline 
prfstf & 0.6227141892 & 0.5851857493 & 0.6309119265\tabularnewline
\hline 
nprfstf & 0.5816411688 & 0.5341568057 & 0.6264055044\tabularnewline
\hline 
studast & 0.2988348852 & 0.2828011591 & 0.3050653567\tabularnewline
\hline 
totstf & 0.6527079425 & 0.6050956364 & 0.6760714085\tabularnewline
\hline 
totstfx & 0.5823181777 & 0.5312762344 & 0.6018386529\tabularnewline
\hline 
explm & 0.5662421434 & 0.5818881361 & 0.5912554113\tabularnewline
\hline 
salprf & 0.6249410208 & 0.6039941805 & 0.6394557823\tabularnewline
\hline 
salnprf & 0.584824465 & 0.5448070294 & 0.6352628324\tabularnewline
\hline 
salstud & 0.3855671652 & 0.3971957008 & 0.3657513915\tabularnewline
\hline 
totsal & 0.6516555093 & 0.6225347339 & 0.6798021027\tabularnewline
\hline 
opexp & 0.5831339988 & 0.5512249133 & 0.5996413111\tabularnewline
\hline 
totexp & 0.6487164033 & 0.6346574034 & 0.669598021\tabularnewline
\hline 
totstu & 0.303291407 & 0.2959357555 & 0.2869635127\tabularnewline
\hline 
totpt & -0.0081385702 & 0.0178274552 & -0.0101793445\tabularnewline
\hline 
gradstu & 0.6810465694 & 0.6703123137 & 0.7022016079\tabularnewline
\hline 
gradpt & 0.0924249587 & 0.0948420614 & 0.1098206555\tabularnewline
\hline 
phdawd & 0.549965402 & 0.4706506373 & 0.5672859281\tabularnewline
\hline 
phdfld & 0.5503286746 & 0.4828899406 & 0.5447268809\tabularnewline
\hline 
fac & 0.5337936162 & 0.5398186804 & 0.5552848193\tabularnewline
\hline 
expbibue & 0.1000711344 & 0.0705293745 & 0.078584155\tabularnewline
\hline 
title & 0.5214960154 & 0.453530459 & 0.5699319728\tabularnewline
\hline 
ebooks & 0.3064482246 & 0.3387216479 & 0.3232776747\tabularnewline
\hline 
exponetime & 0.4729862207 & 0.5159281474 & 0.4809847897\tabularnewline
\hline 
expongoing & 0.5359050549 & 0.5095102437 & 0.5516219902\tabularnewline
\hline 
expcollsup & 0.22845118 & 0.2267792546 & 0.2288453217\tabularnewline
\hline 
fulltextarticlerequests & 0.4558011687 & 0.408169935 & 0.479992685\tabularnewline
\hline 
regsearches & 0.1060937816 & 0.0985185015 & 0.1088300447\tabularnewline
\hline 
fedsearches & 0.0123156003 & 0.0480469033 & 0.0323813389\tabularnewline
\hline 
RESSPACE & 0.7685969811 & 0.6146944551 & 0.7830475142\tabularnewline
\hline 
RESSPACENEW & 0.1246734965 & 0.1828950761 & 0.1465507821\tabularnewline
\hline 
\end{tabular}
\end{table}

\begin{table}[h]

\protect\caption{Correlations (Spearman's rho) between RDCAT, RDBIN, and RD, continued}

\begin{tabular}{|c|c|c|c|}
\hline 
 & RDCAT & RDBIN & RD\tabularnewline
\hline 
\hline 
LibrarianTenure & -0.0512525736 & -0.0329449774 & -0.0200976469\tabularnewline
\hline 
Researchtotalexp & 0.9105849852 & 0.8129319549 & 0.9491280148\tabularnewline
\hline 
Researchsalaries & 0.9181024372 & 0.8115057585 & 0.9525541126\tabularnewline
\hline 
Researchfringebenefits & 0.8849938074 & 0.7872628538 & 0.920305876\tabularnewline
\hline 
Researchplantmaintops & 0.6971850576 & 0.6225424339 & 0.748320944\tabularnewline
\hline 
Researchdepreciation & 0.7619328156 & 0.7002624383 & 0.8043661101\tabularnewline
\hline 
Researchinterest & 0.4726724887 & 0.3689228259 & 0.4828765275\tabularnewline
\hline 
Researchother & 0.8563427914 & 0.7808425356 & 0.8956091528\tabularnewline
\hline 
Endowment & 0.5127811499 & 0.5497987169 & 0.5357823129\tabularnewline
\hline 
AAU & 0.6416256311 & 0.6034053156 & 0.661042037\tabularnewline
\hline 
InstControl & -0.0512941955 & -0.1114064093 & -0.0617344098\tabularnewline
\hline 
Hospital & 0.1230363617 & 0.0429989469 & 0.1392101775\tabularnewline
\hline 
MedicalDegree & 0.3611610595 & 0.2620635907 & 0.3658730159\tabularnewline
\hline 
LandGrant & 0.1440141434 & 0.1152780835 & 0.1342151924\tabularnewline
\hline 
Masters & 0.492596019 & 0.5045247812 & 0.5113870036\tabularnewline
\hline 
PHDResearch & 0.7836176143 & 0.6931400295 & 0.8054706923\tabularnewline
\hline 
PHDProfPractice & 0.3247344129 & 0.2870744056 & 0.3460309762\tabularnewline
\hline 
FTNoninsstaffno & 0.7119990299 & 0.6328766148 & 0.739958998\tabularnewline
\hline 
FTNoninsstaffexp & 0.7378692854 & 0.6688861173 & 0.7683487941\tabularnewline
\hline 
LibCurArchotherno & 0.4650582831 & 0.4164557911 & 0.448881399\tabularnewline
\hline 
LibCurArchotherexp & 0.503169825 & 0.4563828519 & 0.498008658\tabularnewline
\hline 
Managementno & 0.2939384652 & 0.2570766728 & 0.3413359556\tabularnewline
\hline 
Managementexp & 0.4122048144 & 0.3779420492 & 0.4630179344\tabularnewline
\hline 
BusFinOpsno & 0.5922102008 & 0.5790446958 & 0.5945175405\tabularnewline
\hline 
BusFinOpsexp & 0.6185953693 & 0.6182561446 & 0.6181818182\tabularnewline
\hline 
CompEngScino & 0.6358132868 & 0.574049381 & 0.6723933741\tabularnewline
\hline 
CompEngSciexp & 0.6951337877 & 0.6503455639 & 0.731886209\tabularnewline
\hline 
CommServLegalArtsMediano & 0.4175081258 & 0.3512128127 & 0.3912439819\tabularnewline
\hline 
CommServLegalArtsMediaexp & 0.4577559696 & 0.4143113388 & 0.4380161967\tabularnewline
\hline 
Healthcareno & 0.4651507602 & 0.362997294 & 0.4679116702\tabularnewline
\hline 
Healthcareexp & 0.507293538 & 0.4100314685 & 0.5119233148\tabularnewline
\hline 
Serviceno & 0.4874173791 & 0.4817037962 & 0.5040322082\tabularnewline
\hline 
Serviceexp & 0.531875734 & 0.5120045119 & 0.5353123067\tabularnewline
\hline 
Salesno & -0.0685108102 & -0.0051171511 & -0.0936441676\tabularnewline
\hline 
Salesexp & -0.0503669937 & 0.0043852461 & -0.0777981866\tabularnewline
\hline 
OfficeAdminno & 0.5271968007 & 0.4578217893 & 0.5579437896\tabularnewline
\hline 
OfficeAdminexp & 0.569719125 & 0.5091521191 & 0.5932467532\tabularnewline
\hline 
NatResourcesConstrMaintno & 0.4353582068 & 0.3647677795 & 0.4303490592\tabularnewline
\hline 
NatResourcesConstrMaintexp & 0.5436641744 & 0.4813412891 & 0.5396165739\tabularnewline
\hline 
ProdTransMatsno & 0.1862931998 & 0.1840704273 & 0.2187291445\tabularnewline
\hline 
ProdTransMatsexp & 0.2034116812 & 0.1890411828 & 0.2323806149\tabularnewline
\hline 
TotalFTEstaff & 0.8092082396 & 0.7259339738 & 0.8413729128\tabularnewline
\hline 
TeachersFTEstaff & 0.7988462679 & 0.7245122579 & 0.8241547567\tabularnewline
\hline 
LibcurarchteachingotherinstrsupportFTE & 0.4589170781 & 0.3997001961 & 0.4427183026\tabularnewline
\hline 
\end{tabular}
\end{table}

\begin{table}[h]

\protect\caption{Correlations (Spearman's rho) between RDCAT, RDBIN, and RD, continued}

\begin{tabular}{|c|c|c|c|}
\hline 
 & RDCAT & RDBIN & RD\tabularnewline
\hline 
\hline 
LibrCurArchFTE & 0.5052655179 & 0.480681702 & 0.5266572688\tabularnewline
\hline 
teachingotherinstrsupportFTE & 0.3056217666 & 0.2375285491 & 0.2739175325\tabularnewline
\hline 
MgmtFTE & 0.3528027382 & 0.3130539845 & 0.3998812602\tabularnewline
\hline 
BusFinOpsFTE & 0.6697807468 & 0.6610522574 & 0.675149585\tabularnewline
\hline 
CompEngSciFTE & 0.7512911286 & 0.6892158096 & 0.7940395242\tabularnewline
\hline 
CommServLegalArtsMediaFTE & 0.460840251 & 0.3747481717 & 0.4390219771\tabularnewline
\hline 
HealthcareFTE & 0.4835030079 & 0.3865075937 & 0.5071785723\tabularnewline
\hline 
AllService...FTE & 0.646413221 & 0.5804637346 & 0.6712327496\tabularnewline
\hline 
ServiceFTE & 0.5725941251 & 0.5394738063 & 0.5816612091\tabularnewline
\hline 
SalesFTE & -0.0725110512 & -0.0040046314 & -0.1038296878\tabularnewline
\hline 
OfficeAdminFTE & 0.5989959315 & 0.5291254135 & 0.6295748866\tabularnewline
\hline 
NatResourceConstMaintFTE & 0.5249189714 & 0.439292946 & 0.5236098492\tabularnewline
\hline 
ProdTransMovingFTE & 0.2304365029 & 0.223669742 & 0.2460122333\tabularnewline
\hline 
\end{tabular}

\end{table}

The correlation matrix reveals very little difference between the
correlations of the categorical versions of research funding versus
the continuous. This is reassuring and supports the idea that the
categorical data is a useful simplification of the data that does
not distort the results. There is never more than a 0.052 difference
in absolute value between the \texttt{RD} and \texttt{RDCAT} correlations.
This indicates that our modeling results should be robust with respect
to the choice of response variable. 

Among the library variables, we eliminate two from consideration on
logical grounds. Region and membership year are not under the control
of the institution, so even if correlations are discovered, they cannot
guide policy. It is also difficult to interpret region in a sensible
manner as an input to grant funding. Membership year (the date the
institution joined ARL) is correlated with research, with an earlier
date implying higher research funding, but this is likely due to its
correlation with the size, prestige, endowment, and other such attributes
of the older schools.

We also eliminate from consideration variables with very low correlations
with research, taking an absolute value of less than 0.1 as the cutoff.
We do not wish to remove too many variables at this stage, in case
they play a secondary role as modifiers in a multivariate model. Among
the library variables, the number of part-time students, part-time
graduate students, federated searches of library databases, and regular
searches of library databases are not correlated. We may conclude
that it is full-time, but not part-time students, that are correlated
with high levels of research. Measures of searching may be of interest
to librarians studying changing modes of access, but they appear to
not be of primary relevance to research output.

After these modifications, there are 29 library-related variables
retained in the dataset, listed in Table 4.

\begin{table}[h]

\protect\caption{Library-related variables from the ARL Statistics}

\begin{tabular}{|c|c|}
\hline 
Abbreviation & Description\tabularnewline
\hline 
\hline 
vols & volumes in Library\tabularnewline
\hline 
illtot & titles loaned to other libraries\tabularnewline
\hline 
ilbtot & titles borrowed from other libraries\tabularnewline
\hline 
grppres & group presentations\tabularnewline
\hline 
presptcp & presentation participants\tabularnewline
\hline 
reftrans & reference transactions\tabularnewline
\hline 
initcirc & initial circulation of books (not counting renewals)\tabularnewline
\hline 
prfstf & professional staff (librarians)\tabularnewline
\hline 
nprfstf & non-professional staff (support)\tabularnewline
\hline 
studast & student assistants\tabularnewline
\hline 
totstf & total staff (librarians+support)\tabularnewline
\hline 
totstfx & total staff (inc. students)\tabularnewline
\hline 
explm & total materials expenditures\tabularnewline
\hline 
salprf & professional salaries\tabularnewline
\hline 
salnprf & non-professional salaries\tabularnewline
\hline 
salstud & student salaries\tabularnewline
\hline 
totsal & total salaries\tabularnewline
\hline 
opexp & operating expenditures\tabularnewline
\hline 
totexp & total expenditures\tabularnewline
\hline 
totstu & total students (at University)\tabularnewline
\hline 
gradstu & graduate students (at University)\tabularnewline
\hline 
phdawd & PhDs awarded\tabularnewline
\hline 
phdfld & fields of PhD study\tabularnewline
\hline 
fac & total teaching faculty\tabularnewline
\hline 
title & number of unique titles held by library\tabularnewline
\hline 
ebooks & ebooks\tabularnewline
\hline 
exponetime & one-time resource expenditures\tabularnewline
\hline 
expongoing & ongoing resource expenditures\tabularnewline
\hline 
expcollsup & collection support expenditures\tabularnewline
\hline 
\end{tabular}

\end{table}

\begin{figure}[h]
\protect\caption{\label{fig:illtot-before-and}illtot before and after log transform}

\includegraphics[width=0.5\columnwidth]{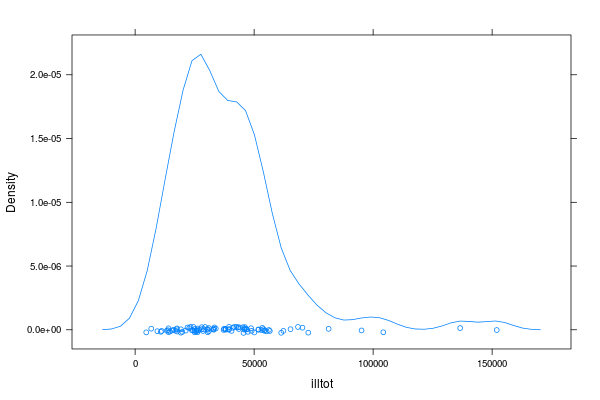}\includegraphics[width=0.5\columnwidth]{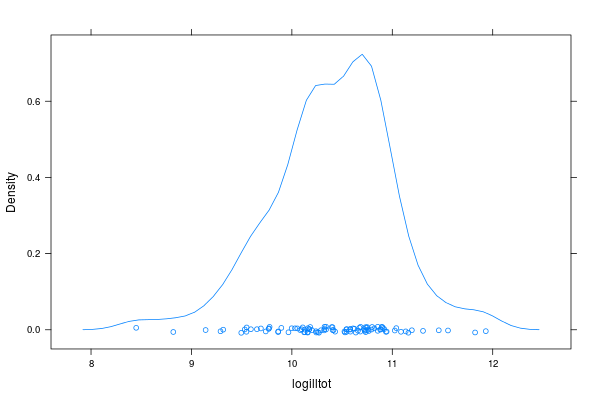}
\end{figure}

Plotting the variables individually reveals that they all have an
asymmetric long-tailed distribution. This is understandable since
these are counts bounded below by zero, and typically a handful of
institutions will have much larger collections, budgets, or staffs
than the mid-size institutions. Using the log transform on all of
the explanatory variables improves their distributions to a more normal
shape. For example, see Figure \ref{fig:illtot-before-and} for the
effect of the log transform on illtot, the number of interlibrary
loan borrowings.

Some skewness and outliers remain in a few cases, but the long tails
are eliminated by the log transform. As a result, we use the log transformed
versions of all continuous variables for all modeling. Throughout
the paper, ``\texttt{log}'' prefixed to the original variable name
indicates the log transformed version of the original data (e.g.,
\texttt{illtot} becomes \texttt{logilltot}).

Next, we run individual linear regressions of all explanatory variables
against nominal research funding (\texttt{RD}). All of the remaining
variables are significant at at least the 0.1 level. While this may
be because there is a strong association of all variables with the
size of the institution, we will keep all variables under consideration
for the next phase of modeling.

\subsection{Modeling Process}

With such a large set of variables, we turn to stepwise regression
to partially automate the variable selection process. In general,
we will examine the results of working from a minimal model (starting
from a common variable like the number of library volumes) and adding
variables, and compare this with the results of a ``maximal'' model
that includes all of the explanatory variables and attempts to drop
them one-by-one. We always allow variables to move in and out of the
model, selecting in both forward and backwards directions, due to
the large number of variable candidates. The Akaike Information Criterion
(AIC) allows us to simultaneously consider the fit and parsimony of
the model. The MASS library in R implements stepwise AIC selection
with the \texttt{stepAIC} function, which we use as the basis of the
variable selection process. The stepAIC process requires cases with
any missing data to be dropped, but once the model is selected, we
re-run the regression with all cases included so that we can report
a complete explanatory fit. 

Sometimes the minimal and maximal starting points converge on the
same final model, but often they do not. In this case, we must use
our judgment to select one. Once a model is selected, we continue
to drop variables whose coefficients are individually insignificant,
as long as AIC does not change by a large amount. There is some judgment
involved, as we are more interested in simplifying the model (and
possibly sacrificing the optimum AIC value) when the number of variables
included is very large. Our primary focus is on the impact of the
individual variables, since we are more interested in how variables
are related to and potentially explain variation in research funding
than in developing the most accurate (in terms of explaining all variation)
or the best predictive model. Therefore, inclusion in or exclusion
from the model is more important for the present analysis than a close
examination of the effect sizes generated by the regression coefficients.

\section{Results}

\subsection{Library Models for Nominal Research Output}

We first present the results of the regression models fit to nominal
research output, using variables from the ARL data set only. These
include descriptive measures of the size of the institution (e.g.,
number of graduate students, PhD's awarded, etc.), but no research
inputs other than library variables.

\subsubsection{Library linear model}

Linear regression model selection, fitting, and diagnostics in this
paper use standard techniques such as those found in Montgomery et.
al. \citep{Montgomery2006}. The linear regression model fit by stepAIC
is
\begin{itemize}
\item $logRD=-4.62+0.252*logilltot-2.16*logexplm-0.57*logsalstud+2.28*logtotexp-0.43*logtotstu+0.38*loggradstu+0.36*logphdawd+0.26*logphdfld+0.94*logexpongoing+0.11*logebooks$
{[}0.6994 adjusted $R^{2}${]}
\end{itemize}
Dropping the e-book variable which is not individually significant,
we get
\begin{itemize}
\item $logRD=-4.54+0.41*loggradstu-0.58*logsalstud+2.29*logtotexp+0.33*logphdawd+0.28*logilltot-0.42*logtotstu+0.32*logphdfld-2.04*logexplm+0.85*logexpongoing$
{[}0.694 adjusted $R^{2}${]}
\end{itemize}
In this equation, the coefficients on the individual variables are
all significant at at least the 0.1 level, and the magnitudes are
quite similar to the original stepAIC model.

\begin{figure}[h]

\protect\caption{\label{fig:Diagnostic-Plots-for}Diagnostic Plots for Library Linear
Model including all significant variables}

\includegraphics[width=1\columnwidth]{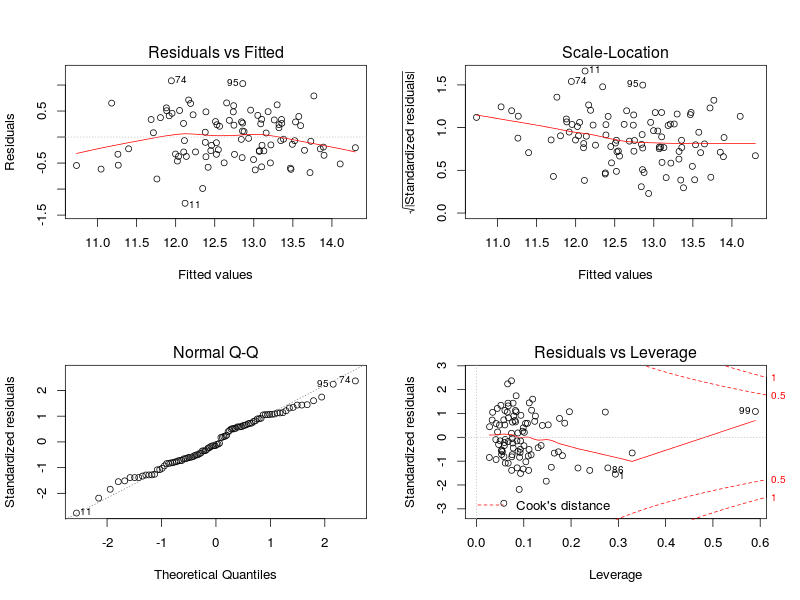}
\end{figure}

\begin{figure}[h]

\protect\caption{\label{fig:Diagnostic-Plots-for-Lvaronly}Diagnostic Plots for Library
Linear Model, library variables only}

\includegraphics[width=1\columnwidth]{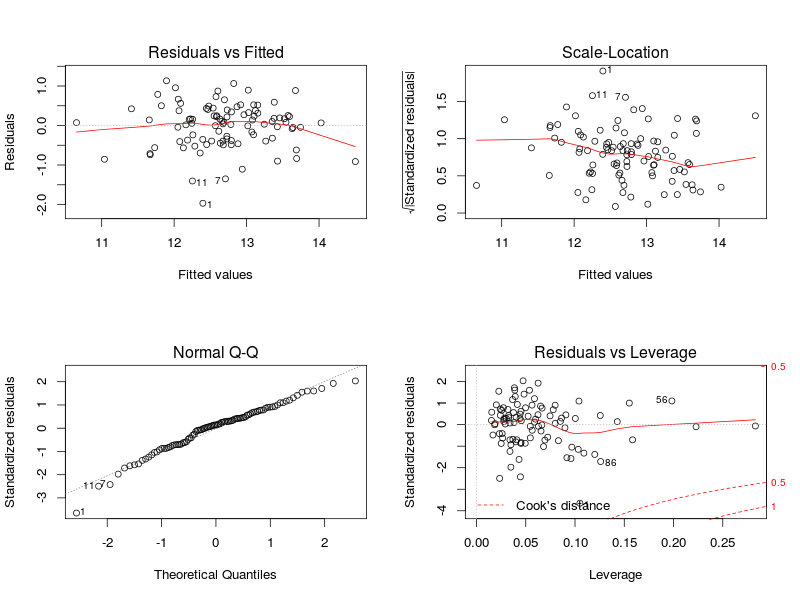}
\end{figure}

We can also consider only variables under library control to isolate
those effects, giving 
\begin{itemize}
\item $logRD=-13.00-0.62*logsalstud+2.90*logtotexp+0.27*logilltot-2.52*logexplm+1.54*logexpongoing$
{[}0.553 adjusted $R^{2}${]}
\end{itemize}
In this case the adjusted $R^{2}$ drops off. Total library expenses,
expenses on ongoing resources, and interlibrary loans are positively
correlated with research funding. These variables relate to the current
strength of the library and its collections. Student salaries and
materials expenditures are negatively related. This is somewhat counterintuitive,
but one possible interpretation is that once the overall spending
and ongoing resources are accounted for, spending on student salaries
(as opposed to professional salaries) and materials expenditures that
are not subscription-based may be associated with less research-oriented
activity. The non-library variables show positive relationships for
PhD's awarded, PhD fields of study, and graduate students, all clearly
associated with research activity. Total students is negatively correlated,
which could be explained as a correcting factor for undergraduate-heavy
institutions. As shown in Figures \ref{fig:Diagnostic-Plots-for}
and \ref{fig:Diagnostic-Plots-for-Lvaronly}, the diagnostic plots
on these regressions show a good fit in general, with only a few outliers
which are the high research institutions at the very end of the scale,
like Johns Hopkins.

To understand the impact of the effects in the library-only model,
consider a change from the 1st quartile to the 3rd quartile of the
range of each of the library variables (in original values, not log
transform), with predicted effect shown below:

\medskip{}

\begin{center}
\begin{tabular}{|c|c|c|c|}
\hline 
 & 1st Quartile & 3rd Quartile & proportional predicted change in Research Funding\tabularnewline
\hline 
\hline 
salstud & 525300 & 1090000 & 0.636\tabularnewline
\hline 
totexp & 18510000 & 34050000 & 5.857\tabularnewline
\hline 
illtot & 24390 & 47190 & 1.20\tabularnewline
\hline 
explm & 8624000 & 12500000 & 0.260\tabularnewline
\hline 
expongoing & 6561000 & 10480000 & 2.057\tabularnewline
\hline 
\end{tabular}
\par\end{center}

\medskip{}

For example, this shows that, according to the model, an increase
in library total expenditure (\texttt{totexp}) from \$18 million to
\$34 million would be expected to be associated with an almost six-fold
increase in research funding. 

We can see that total library expenditure has a much stronger positive
relationship with research funding than other variables, but that
this is counteracted by materials expenditure (\texttt{explm}). If
materials expenditure increased with all other things equal (implying
that total expenditure remained constant), the model predicts a reduction
in research funding. In reality, all of these variables are linked
and would change simultaneously, so we are looking at relative effects
that must be interpreted in context.

\subsubsection{Library binary model}

Next we consider the model with a binary High/Low research level as
the outcome. The binary outcome is modeled with standard logistic
regression techniques such as those in Agresti \citep{Agresti2012}.
A logit link function is used, so the model equation predicts the
logit, or log odds ratio of being in the High category. With the large
number of variables, some manual intervention and selective dropping
of insignificant variables is required in order to achieve convergence
and successful stepwise AIC from the minimal model starting point.
The final selected model, with AIC of 64.83, is:
\begin{itemize}
\item $logit(RDBIN)=-79.45+5.80*loggradstu-3.44*logstudast+2.01*logexponetime+2.92*logprfstf$ 
\end{itemize}
Deviance and deviance residuals indicate a good fit. This model provides
a simple explanation of research funding as a function of graduate
students, student assistants, one-time library expenditures, and library
professional staff. Professional staff is significant at the 0.1 level,
while all other variables are significant at 0.05 or higher. Attempting
to remove \texttt{loggradstu} worsens AIC significantly, so we do
not simplify the model further than this. Only student assistants
have a negative relationship to research, presumably because they
are less closely related to research activity and may substitute for
professional employment.

Note that the magnitudes of the staffing variables are relatively
small. Moving from the 1st quartile of professional staffing (\texttt{logprfstf})
at 62 to the 3rd quartile at 116.5 multiplies the odds of being in
the high research category by 1.84 times.

\subsubsection{Library clm model}

We use a cumulative link model (clm) with logit link to model 4-category
ordinal data. The cumulative logit model allows different threshold
probabilities for each of the four categories, but provides proportional
odds ratios on the predictors, which are easy to interpret. The documentation
for the \texttt{ordinal} package in R contains an excellent outline
on the use and interpretation of cumulative link models \citep{ordinal}.
Agresti also discusses clm models. Like logisitic regression, these
models generate an equation that predicts the logit function or log
odds of the probability of being in the category of interest. The
exact functional form of the cumulative logit differs for each category
level, with a different intercept term for each level. Since we are
more interested in the overall effect of each explanatory variable
rather than the category-level predictions, we report the coefficients
on each variable with the separate intercepts for each category omitted.
The odds are proportional for each category in the cumulative logit. 

In this case, the stepAIC function does not converge from the maximal
model including all variables. But starting from a minimal model (using
\texttt{logvols} as the starting variable) achieves convergence. Individually
dropping insignificant variables from this model gives the following
final model, where the logit of the probability of being in a particular
research category is proportional to: 
\begin{itemize}
\item $logit(RDCAT)\approx1.49*loggradstu+0.99*logilltot-2.39*logstudast+4.05*logtotstfx+2.08*logphdfld$
\end{itemize}
All variables are significant at the 0.05 level. Again, deviance and
deviance residuals indicate a good fit. The variables selected are
similar in nature to the binary case, although, interestingly, there
are no expenditure variables in this model. The size of the research
program of the institution is represented by a positive relationship
with graduate students and PhD fields, while library \textquotedblleft intensity\textquotedblright{}
is represented by a positive relationship with interlibrary lending
and total staffing including students, along with a negative relationship
with the number of student assistants. Using categorical and ordinal
representations of the data has resulted in a simpler model which
is perhaps easier to interpret, compared to the linear model.

\subsection{Academic Models for Nominal Research Output}

We now build models based on the other academic variables from IPEDS
and NCSES. The only variables that are dropped from modeling on the
basis of low univariate correlation are the presence of a tenure system
for librarians, the number of Sales employees, Sales FTEs, and expenditure
on Sales employees. Apparently Sales is one of the support staff categories
unrelated to research output.

Institutional control (public or private) is also not strongly correlated
with research, but the public/private status of a university has a
major influence on the nature of the organization in other ways, so
we leave it in to see if it will enter a model at a later stage. 

The variables related to research itself (Research expense, Research
salaries, etc.) are highly correlated with research at 0.85 or greater,
but there is an endogeneity problem. The research grants themselves
directly fund many of the salaries and expenses of research, so we
choose to drop these from the model. To some extent, the same argument
could be made for research and laboratory space, which in many cases
would be built from previous research grants. However, there is more
potential for a university to construct this space on its own, and
in any case it results from prior grants, not the current research
cycle, so we leave these variables in. 

There are still 51 explanatory variables remaining, so we will have
some work to do in selecting our models. Many variables are slightly
different ways of measuring the same thing, such as the number of
Service staff, the number of FTE Service staff, and the expense on
Service staff. The complete list of academic variables is listed in
Tables 5 and 6, which also include the research variables used throughout
the study.

\begin{table}[h]

\protect\caption{Academic variables from NCSES, IPEDS}

\begin{tabular}{|c|c|c|}
\hline 
{\footnotesize{}abbreviation} & {\footnotesize{}description} & {\footnotesize{}source}\tabularnewline
\hline 
\hline 
{\footnotesize{}RD} & {\footnotesize{}total research funding awarded} & {\footnotesize{}HERD}\tabularnewline
\hline 
{\footnotesize{}RDCAT} & {\footnotesize{}4-category research rank} & {\footnotesize{}constructed}\tabularnewline
\hline 
{\footnotesize{}RDBIN} & {\footnotesize{}2-category research rank} & {\footnotesize{}constructed}\tabularnewline
\hline 
{\footnotesize{}Rdpc} & {\footnotesize{}research funding per faculty member} & {\footnotesize{}constructed}\tabularnewline
\hline 
{\footnotesize{}RDCATpc} & {\footnotesize{}4-category rank based on per capita data} & {\footnotesize{}constructed}\tabularnewline
\hline 
{\footnotesize{}RDBINpc} & {\footnotesize{}2-category rank based on per capita data} & {\footnotesize{}constructed}\tabularnewline
\hline 
{\footnotesize{}RESSPACE} & {\footnotesize{}research space} & {\footnotesize{}SSEF}\tabularnewline
\hline 
{\footnotesize{}RESSPACENEW} & {\footnotesize{}newly constructed research space} & {\footnotesize{}SSEF}\tabularnewline
\hline 
{\footnotesize{}Endowment} & {\footnotesize{}value of endowment} & {\footnotesize{}IPEDS}\tabularnewline
\hline 
{\footnotesize{}AAU} & {\footnotesize{}member of AAU (Yes=1)} & {\footnotesize{}IPEDS}\tabularnewline
\hline 
{\footnotesize{}InstControl} & {\footnotesize{}public or private (Public=1)} & {\footnotesize{}IPEDS}\tabularnewline
\hline 
{\footnotesize{}Hospital} & {\footnotesize{}hospital at university (Yes=1)} & {\footnotesize{}IPEDS}\tabularnewline
\hline 
{\footnotesize{}MedicalDegree} & {\footnotesize{}medical degree granted (Yes=1)} & {\footnotesize{}IPEDS}\tabularnewline
\hline 
{\footnotesize{}LandGrant} & {\footnotesize{}land-grant university (Yes=1)} & {\footnotesize{}IPEDS}\tabularnewline
\hline 
{\footnotesize{}Masters} & {\footnotesize{}number of Master's granted} & {\footnotesize{}IPEDS}\tabularnewline
\hline 
{\footnotesize{}PHDResearch} & {\footnotesize{}number of research PhD's granted} & {\footnotesize{}IPEDS}\tabularnewline
\hline 
{\footnotesize{}PHDProfPractice} & {\footnotesize{}number of PhD's of professional practice} & {\footnotesize{}IPEDS}\tabularnewline
\hline 
{\footnotesize{}FTNoninsstaffno} & {\footnotesize{}full-time non-instructional staff, number} & {\footnotesize{}IPEDS}\tabularnewline
\hline 
{\footnotesize{}FTNoninsstaffexp} & {\footnotesize{}full-time non-instructional staff, expense} & {\footnotesize{}IPEDS}\tabularnewline
\hline 
{\footnotesize{}LibCurArchotherno} & {\footnotesize{}librarians, curators, and archivists, number} & {\footnotesize{}IPEDS}\tabularnewline
\hline 
{\footnotesize{}LibCurArchotherexp} & {\footnotesize{}librarians, curators, and archivists, expense} & {\footnotesize{}IPEDS}\tabularnewline
\hline 
{\footnotesize{}Managementno} & {\footnotesize{}management staff, number} & {\footnotesize{}IPEDS}\tabularnewline
\hline 
{\footnotesize{}Managementexp} & {\footnotesize{}management staff, expense} & {\footnotesize{}IPEDS}\tabularnewline
\hline 
{\footnotesize{}BusFinOpsno} & {\footnotesize{}business, finance, and operations staff, number} & {\footnotesize{}IPEDS}\tabularnewline
\hline 
{\footnotesize{}BusFinOpsexp} & {\footnotesize{}business, finance, and operations staff, expense} & {\footnotesize{}IPEDS}\tabularnewline
\hline 
{\footnotesize{}CompEngScino} & {\footnotesize{}computing, engineering, and scientific staff, number} & {\footnotesize{}IPEDS}\tabularnewline
\hline 
{\footnotesize{}CompEngSciexp} & {\footnotesize{}computing, engineering, and scientific staff, expense} & {\footnotesize{}IPEDS}\tabularnewline
\hline 
{\footnotesize{}CommServLegalArtsMediano} & {\footnotesize{}communication services, legal, arts, media staff,
number} & {\footnotesize{}IPEDS}\tabularnewline
\hline 
{\footnotesize{}CommServLegalArtsMediaexp} & {\footnotesize{}communication services, legal, arts, media staff,
expense} & {\footnotesize{}IPEDS}\tabularnewline
\hline 
\end{tabular}
\end{table}

\begin{table}[h]

\protect\caption{Academic variables from NCSES, IPEDS, continued}

\begin{tabular}{|c|c|c|}
\hline 
{\footnotesize{}abbreviation} & {\footnotesize{}description} & {\footnotesize{}source}\tabularnewline
\hline 
\hline 
{\footnotesize{}Healthcareno} & {\footnotesize{}healthcare staff, number} & {\footnotesize{}IPEDS}\tabularnewline
\hline 
{\footnotesize{}Healthcareexp} & {\footnotesize{}healthcare staff, expense} & {\footnotesize{}IPEDS}\tabularnewline
\hline 
{\footnotesize{}Serviceno} & {\footnotesize{}service staff, number} & {\footnotesize{}IPEDS}\tabularnewline
\hline 
{\footnotesize{}Serviceexp} & {\footnotesize{}service staff, expense} & {\footnotesize{}IPEDS}\tabularnewline
\hline 
{\footnotesize{}OfficeAdminno} & {\footnotesize{}office administrative staff, number} & {\footnotesize{}IPEDS}\tabularnewline
\hline 
{\footnotesize{}OfficeAdminexp} & {\footnotesize{}office administrative staff, expense} & {\footnotesize{}IPEDS}\tabularnewline
\hline 
{\footnotesize{}NatResourcesConstrMaintno} & {\footnotesize{}natural resources, construction, and maintenance,
number} & {\footnotesize{}IPEDS}\tabularnewline
\hline 
{\footnotesize{}NatResourcesConstrMaintexp} & {\footnotesize{}natural resources, construction, and maintenance,
expense} & {\footnotesize{}IPEDS}\tabularnewline
\hline 
{\footnotesize{}ProdTransMatsno} & {\footnotesize{}production, transportation, and moving, number} & {\footnotesize{}IPEDS}\tabularnewline
\hline 
{\footnotesize{}ProdTransMatsexp} & {\footnotesize{}production, transportation, and moving, expense} & {\footnotesize{}IPEDS}\tabularnewline
\hline 
{\footnotesize{}TotalFTEstaff} & {\footnotesize{}total staff in FTE (full-time equivalent)} & {\footnotesize{}IPEDS}\tabularnewline
\hline 
{\footnotesize{}TeachersFTEstaff} & {\footnotesize{}total teachers, FTE} & {\footnotesize{}IPEDS}\tabularnewline
\hline 
{\scriptsize{}LibcurarchteachingotherinstrsupportFTE} & {\footnotesize{}librarians, curators, and archivists, number} & {\footnotesize{}IPEDS}\tabularnewline
\hline 
{\footnotesize{}LibrCurArchFTE} & {\footnotesize{}librarians, curators, and archivists, FTE} & {\footnotesize{}IPEDS}\tabularnewline
\hline 
{\footnotesize{}teachingotherinstrsupportFTE} & {\footnotesize{}librarians, curators, and archivists, number} & {\footnotesize{}IPEDS}\tabularnewline
\hline 
{\footnotesize{}MgmtFTE} & {\footnotesize{}management staff, FTE} & {\footnotesize{}IPEDS}\tabularnewline
\hline 
{\footnotesize{}BusFinOpsFTE} & {\footnotesize{}business, finance, and operations staff, FTE} & {\footnotesize{}IPEDS}\tabularnewline
\hline 
{\footnotesize{}CompEngSciFTE} & {\footnotesize{}computing, engineering, and scientific staff, FTE} & {\footnotesize{}IPEDS}\tabularnewline
\hline 
{\footnotesize{}CommServLegalArtsMediaFTE} & {\footnotesize{}communication services, legal, arts, media staff,
FTE} & {\footnotesize{}IPEDS}\tabularnewline
\hline 
{\footnotesize{}HealthcareFTE} & {\footnotesize{}healthcare staff, FTE} & {\footnotesize{}IPEDS}\tabularnewline
\hline 
{\scriptsize{}AllServiceinclsalesofficeadminconstrmaintprodtransFTE} & {\footnotesize{}all service categories combined, FTE} & {\footnotesize{}IPEDS}\tabularnewline
\hline 
{\footnotesize{}ServiceFTE} & {\footnotesize{}service staff, FTE} & {\footnotesize{}IPEDS}\tabularnewline
\hline 
{\footnotesize{}OfficeAdminFTE} & {\footnotesize{}office administrative staff, FTE} & {\footnotesize{}IPEDS}\tabularnewline
\hline 
{\footnotesize{}NatResourceConstMaintFTE} & {\footnotesize{}natural resources, construction, and maintenance,
FTE} & {\footnotesize{}IPEDS}\tabularnewline
\hline 
{\footnotesize{}ProdTransMovingFTE} & {\footnotesize{}production, transportation, and moving, FTE} & {\footnotesize{}IPEDS}\tabularnewline
\hline 
\end{tabular}

\end{table}

Once again, the continuous variables are bounded below and have longer
upper tails due to a few extreme values. Log transformation is applied
to all continuous variables to restore them to normality.

\subsubsection{Fisher's Exact Test}

As a preliminary step, we will use Fisher's Exact test on the categorical
predictors to see if they are related to the binary classification
of research funding or not. AAU membership and Medical Degree-granting
are significant, while Institutional Control, Land Grant status, and
presence of a Hospital are not. AAU membership has an estimated odds-ratio
of 21.51 (with 95\% confidence interval of 6.38 to 96.07), so it has
a strong effect. Since AAU universities are considered to be the largest
and most research intensive, this result is not surprising. The odd-ratio
of granting a medical degree is smaller at 3.65 (with 95\% confidence
interval of 1.24 to 12.41). Hospitals are not clearly related to research
funding, perhaps because some hospitals affiliated with university
medical schools are not directly under university control.

\subsubsection{Academic linear model}

After running individual regressions against each variable, we drop
the following variables which are not significant at the 0.1 level:
\texttt{logRESSPACENEW}, \texttt{logCommServLegalArtsMediano}, \texttt{}~\\
\texttt{logCommServLegalArtsMediaexp}, \texttt{logProdTransMatsno,}
\texttt{logProdTransMatsexp}, and \texttt{}~\\
\texttt{logProdTransMovingFTE}. We also drop \texttt{logNatResourcesConstrMaintexp}
and \texttt{}~\\
\texttt{logCommServLegalArtsMediaFTE}, which although significant
at 0.1, are similar to the other employment variables in these categories
which have even higher significance.

\begin{figure}[h]
\protect\caption{\label{fig:Diagnostic-Plots-for-Acad-Linear}Diagnostic Plots for
Academic Linear Model, academic variables only}

\includegraphics[width=1\columnwidth]{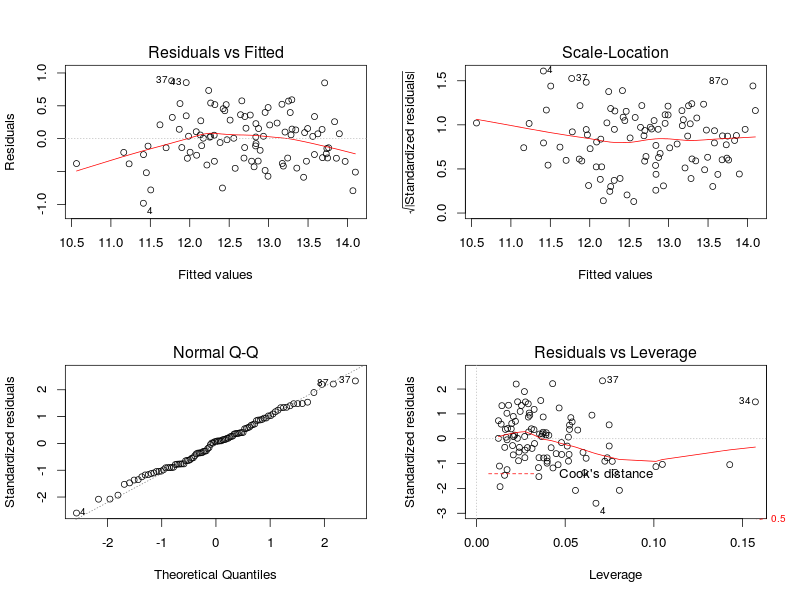}
\end{figure}

After several iterations to deal with the large number of similar
variables, we get the following model
\begin{itemize}
\item $logRD=-0.21+0.55*logRESSPACE+0.47*logCompEngSciFTE+0.11*logEndowment+0.57*logPHDResearch+0.93*logCompEngSciExp+0.49*logBusFinOpsFTE-0.12*logNatResourceConstMaintFTE-1.3*logCompEngScino-0.16*logMgmtFTE-0.52*logBusFinOpsexp$
\end{itemize}
All variables are significant at the 0.01 level, and adjusted $R^{2}$
is 0.8956. Diagnostic plots show a reasonable fit, illustrated in
Figure \ref{fig:Diagnostic-Plots-for-Acad-Linear}. Looking at reduced
models, we can achieve an adjusted $R^{2}$ of 0.7876 with the following
simple equation:
\begin{itemize}
\item $logRD=2.90+0.57*logRESSPACE+0.15*logEndowment+0.48*logPHDResearch$. 
\end{itemize}

The variables in this version of the academic linear model are mostly
related to the size of the institution and its research activity.
Several employment categories are related, and the magnitude of the
effect is much greater for employment in computing, engineering, and
science. Note that CompEngSci has two positive terms (FTE and expenses)
and one negative (number). A speculative interpretation of this result
would be that a high number of CompEngSci employees along with low
expenses and FTEs would indicate a large pool of low-level, part-time
workers and a less intensive scientific research program. But the
majority of the magnitude of research funding can be explained by
considering only research space, endowment, and the number of research
PhD's granted as explanatory factors.

\subsubsection{Academic binary model}

Here we model the binary outcome of high/low research activity using
the non-library academic explanatory variables for each institution.
After studying the results of individual logistic regressions, a similar
selection of non-significant variables is removed before interactive
modeling via stepwise AIC. PhD's in professional practice, FTEs in
teaching and other instructional support, and employment categories
such as communications, legal, arts, media and production, transportation,
and moving do not make the initial cut.

After some tweaking, the stepwise AIC selection process yields the
following simple model:
\begin{itemize}
\item $logit(RDBIN)=-110.25+11.65*logTotalFTEstaff+1.79*logEndowment-4.79*logOfficeAdminFTE-2.09*logMgmtFTE+2.00*logRESSPACE$
\end{itemize}
Deviance and deviance residuals indicate a good fit. All variables
are significant at 0.05 level or greater. It is interesting that this
model uses total FTE staff for its primary positive effective rather
than any specific job category. Overall size of the institution appears
to be the dominant effect. Having too many office administrative or
management staff is associated with lower research output.

\subsubsection{Academic clm model}

As before, the clm model uses a four-category classification of research
output as the response variable, now modeled with the non-library
academic explanatory variables. After the usual tweaking of the stepwise
AIC process, the following model was selected: 
\begin{itemize}
\item $logit(RDCAT)\approx3.38*logRESSPACE+9.79*logTotalFTEstaff+1.03*logEndowment-4.50*logAllServiceinclsalesofficeadminconstrmaintprodtransFTE-1.83*logMgmtFTE$
\end{itemize}
All variables are significant at <0.001 level here. This model is
not significantly different by ANOVA from the model with the lowest
AIC, and has the advantage of being simpler and having tightly defined
parameter coefficients (within narrow CIs). Here the ``All Service''
category of employment replaces office administrative staff in the
logistic model. The coefficients are also roughly similar, but the
effect of research space has increased while endowment effect has
decreased.

\subsection{Combined Models for Nominal Research Output}

We now determine which of the library or academic variables retain
significance in a model that allows each of these groups of variables
to enter. Considering all of the initial variables in the dataset
will not be feasible given the limited number of observations. Instead,
we use the results of our analysis above to develop our pool of variables.
The combined model for each type is generated by including the variables
in the final library equation and the variables in the final academic
equation, then using stepwise AIC with forward and backward inclusion
to generate the final model.

\subsubsection{Combined linear model}

When we fit the linear model, the best fit is generated by the identical
variables as those in the academic-only case. In other words, the
library values do not enter the model, and add no explanatory value.
The equation has slightly different coefficients, presumably as a
result of a slightly different path of iterative estimation. The impact
and interpretation of the variables is the same as before.
\begin{itemize}
\item $logRD=-0.63+0.53*logRESSPACE+0.46*logCompEngSciFTE+0.11*logEndowment+0.51*logPHDResearch+0.95*logCompEngSciExp+0.48*logBusFinOpsFTE-0.12*logNatResourceConstMaintFTE-1.31*logCompEngScino-0.16*logMgmtFTE-0.52*logBusFinOpsexp$
\end{itemize}

\subsubsection{Combined binary model}

The best fit is generated by:
\begin{itemize}
\item $logit(RDBIN)=-105.97+2.32*logRESSPACE+1.02*logEndowment+6.00*loggradstu-2.46*logstudast+2.29*logexponetime-1.18*logMgmtFTE$
\end{itemize}
All variables are significant at the 0.1 level or better. Once again,
deviance and deviance residuals show no lack of fit. The coefficients
are similar to what we have seen in other models. The library variables
for one-time expenses and student assistants enter the model. These
are not the variables that one might expect to have the most impact,
but they appear to explain some of the residual differences after
research space, endowment, and graduate students enter the model.

\subsubsection{Combined clm model}

In this case, our preferred model, which is not significantly different
than the stepAIC-generated four-variable model with FTE Management
staff included, is:
\begin{itemize}
\item $logit(RDCAT)\approx2.92*logRESSPACE+0.73*logEndowment+1.68*loggradstu$
\end{itemize}
No library-specific component enters the model. The graduate student
count from the ARL data renders the other measures in the academic
clm model unnecessary. Here the category of research funding is directly
related to the university's financial resources, physical space for
research, and the size of the graduate program. This is simple and
intuitive, but it also provides no evidence for the impact on research
funding of other inputs to the research process (libraries, computing,
or otherwise).

\subsection{Library Models for Per Capita Output}

As discussed by Hendrix, it is important to analyze size-independent
measures. The amount of research funding is strongly correlated with
all measures of the size of the university, from enrollments and employment
to endowments. Our variables may have entered the nominal models purely
from this kind of correlation. To understand relationships between
inputs and research funding that persist across institutions regardless
of size, we will repeat the same steps of analysis with per capita
measures as the response variable.

In this section, we take research funding per capita, defined as research
funding divided by the number of faculty, as our response variable.
This does make a difference in rankings, as we can see below:

\medskip{}

\begin{center}
\begin{tabular}{|c|c|c|}
\hline 
ranking & Per Capita funding & Total funding\tabularnewline
\hline 
\hline 
1 & Johns Hopkins & Johns Hopkins\tabularnewline
\hline 
2 & UCSD & Michigan\tabularnewline
\hline 
3 & MIT & Wisconsin\tabularnewline
\hline 
4 & Duke & Washington\tabularnewline
\hline 
5 & Case Western Reserve & UCSD\tabularnewline
\hline 
\end{tabular}
\par\end{center}

\medskip{}

As before, for categorical data analysis we define four categories
of activity, outlined below. Here the cutpoints are set to get generate
almost equal numbers in each category. For binary analysis, low is
simply below \$225,000 per faculty, and high is above \$225,000 per
faculty.

\medskip{}

\begin{center}
\begin{tabular}{|c|c|c|}
\hline 
 & Range (in thousands of dollars per faculty) & Number\tabularnewline
\hline 
\hline 
1 & 0-152 & 25\tabularnewline
\hline 
2 & 152-225 & 24\tabularnewline
\hline 
3 & 225-350 & 25\tabularnewline
\hline 
4 & 350- & 25\tabularnewline
\hline 
\end{tabular}
\par\end{center}

\medskip{}

We drop the following variables from consideration for lack of correlation
with per capita research funding: \texttt{InstControl}, \texttt{totstu},
\texttt{fac}, \texttt{LandGrant}, \texttt{presptcp}, \texttt{grppres},
\texttt{reftrans}, \texttt{studast}, \texttt{ProdTransMatsno}, \texttt{ProdTransMovingFTE},
and \texttt{Hospital}. After considering individual regressions, we
drop \texttt{logsalstud} and \texttt{logexpcollsup} for lack of significance.
We also use the log transform of per capita research funding to normalize
its distribution.

\subsubsection{Library linear per capita model}

Our preferred model is:
\begin{itemize}
\item $logRdpc=-9.20+0.28*logphdawd+0.36*logilltot+0.96*logsalprf+0.67*lognprfstf-1.63*logtotstfx$
\end{itemize}
This model has familiar variables representing collection uniqueness
(\texttt{logilltot}) and research activity (\texttt{logphdawd}). It
places considerable emphasis on the level of staffing in the professional
and support ranks of the library, while only total staff including
students is negatively associated with research. All variables are
significant at the 0.01 level. However, adjusted $R^{2}$ is only
0.291, so we are explaining much less of the variation in funding
in the per capita case compared to the nominal case.

\subsubsection{Library binary per capita model}

After the stepwise AIC selection process, the preferred model, which
also has the lowest AIC is 
\begin{itemize}
\item $logit(RDBINpc)=-42.92+1.47*logilltot+2.51*logsalprf-2.09*logtotstfx$
\end{itemize}
The variable \texttt{logtotstfx} is only significant at the 0.07 level,
but since this is already a parsimonious model, we retain it. This
model focuses exclusively on library-specific variables, with similar
relationships to those in the linear model. Professional salaries
is the lone positive variable from the staffing side, while total
staffing including students is negative. This may be interpreted as
a higher percentage of library professional staff and higher paid
library professional staff being associated with more research activity.
The now familiar interlibrary loan total plays a positive role as
well.

Deviance residuals are within normal limits. The residual deviance
indicates that this model explains less variation than the nominal
binary model.

\subsubsection{Library clm per capita model}

For the four category model, the selection process converges quickly
to
\begin{itemize}
\item $logit(RDCATpc)\approx2.01*logsalprf+1.02*logillot+0.65*logphdawd-1.44*logsalstud$
\end{itemize}
The coefficient for \texttt{logphdawd} is significant at the 0.1 level,
while the others are significant at the 0.01 level. The patterns in
the data are similar to the previous two models, with student salaries
taking the place of \texttt{totsftfx} as the negative effect. All
of the per capita variants of the library models include interlibrary
loans and library professional salaries as positive correlates of
research funding.

\subsection{Academic Models for Per Capita Output}

With per capita research output as the response variable, the models
generated from the academic data tend to be more complex than other
models, with many variables included. In contrast to many of the other
cases, additional variables cannot be dropped without large changes
in AIC. We are left with models with many mixtures of effects, as
reflected below. The models are presented briefly below, and will
be discussed further when the models are compared.

\subsubsection{Fisher's exact test, per capita}

The \texttt{AAU} and \texttt{MedicalDegree} indicator variables retain
their significance against per capita measures. AAU membership has
an odds ratio for high research activity of 5.35 (95\% CI is 2.11
to 14.37). Offering a medical degree has odds ratio 3.28 (95\% CI
is 1.18 to 9.90).

\subsubsection{Academic linear per capita model}

\begin{figure}[h]
\protect\caption{\label{fig:Diagnostic-Plots-for-Acad-Linear-pc}Diagnostic Plots for
Academic Linear Model, per capita case}

\includegraphics[width=1\columnwidth]{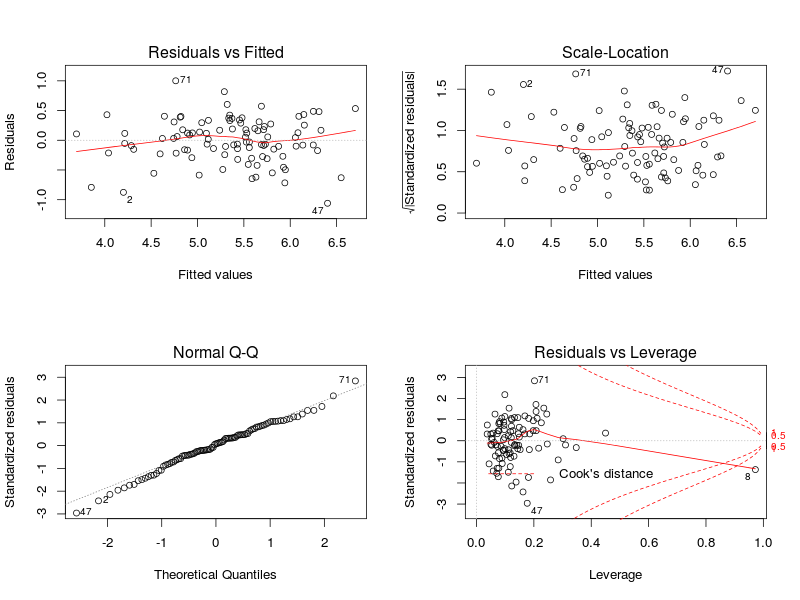}
\end{figure}

The best fitting model, with 13 explanatory variables is the following:
\begin{itemize}
\item $logRdpc=-9.48+0.49*logRESSPACE+0.43*logPHDResearch-1.43*logFTNoninsstaffno+1.90*logFTNoninsstaffexp-0.27*logManagementexp-0.92*logCompEngScino+0.1*logHealthcareno-0.19*logServiceexp-0.80*logTotalFTEstaff+1.06*logCompEngSciFTE-1.90*logAllServiceFTE+0.74*logServiceFTE+1.06*logOfficeAdminFTE$
\end{itemize}
The staffing effects are somewhat complex with positive and negative
coefficients for absolute numbers, FTEs, and expenses in several employment
categories. This model may overfit the data, using several similar
variables to fit small variations in research. As an explanatory model,
it is difficult to interpret. However, adjusted $R^{2}$ is 0.704,
much better than the library linear per capita model. Diagnostic plots,
shown in Figure \ref{fig:Diagnostic-Plots-for-Acad-Linear-pc}, show
this model fits the data well. Note that \texttt{AllServiceFTE} is
an abbreviation for \texttt{AllServiceinclsalesofficeadminconstrmaintprodtransFTE}
in the dataset.

\subsubsection{Academic binary per capita model}

The variables omitted as insignificant after individual regressions
are very similar to those omitted in the nominal case. The selected
model in the binary outcome case reduces deviance by more than the
library binary per capita model, but not as much as the nominal model.
Deviance residuals do not indicate lack of fit.
\begin{itemize}
\item $logit(RDBINpc)=-115.01+2.32*logRESSPACE-9.93*logCompEngScino+9.10*logCompEngSciexp+2.83*logCompEngSciFTE-2.75*logLibcurarchteachingotherinstrsupportFTE+1.17*logteachingotherinstrsupport-0.97*logNatResourceConstrMaintFTE$
\end{itemize}
All parameter coefficients are significant at 0.01 or less, except
for the coefficient of \texttt{logNatResourceConstMaintFTE}, which
is significant at the 0.1 level. The complex fit on CompEngSci staffing
is notable, with expenditure and FTE being positive, while the actual
number of staff is negative. We may hypothesize that a high number
of part-time staff is associated with a less active researcher program.
Other coefficients are similar to previous models.

\subsubsection{Academic clm per capita model}

Our selected model is
\begin{itemize}
\item $logit(RDCATpc)\approx-3.47*logTotalFTEstaff+2.23*logRESSPACE-8.95*logFTNoninsstaffno+8.30*logFTNoninsstaffexp+5.11*logCompEngSciFTE-4.49*logCompEngscino+0.90*logServiceFTE+1.88*logPHDResearch$
\end{itemize}
All coefficients are significant at 0.05 except for \texttt{logServiceFTE},
which is significant at the 0.1 level. Some of these variables are
the same or similar to the academic binary per capita model. Notable
additions are the positive relationship to the number of research
PhD's granted and the negative relationship to total FTE staff.

\subsection{Combined Models for Per Capita Output}

As in the nominal case, we build models for linear, binary, and clm
from the combined pool of variables selected in the library per capita
models and the academic per capita models. In this case, the models
selected are quite easy to describe because they are nearly identical
to the academic per capita models, with one exception.

\subsubsection{Combined linear per capita model}

Coefficients on all variables are significant at at least 0.05, except
for the coefficient of \texttt{logHealthcareno}, which is significant
at the 0.1 level. Adjusted $R^{2}$ is 0.693.
\begin{itemize}
\item $logRdpc=-9.04+0.46*logRESSPACE+0.48*logPHDResearch-1.46*logFTNoninsstaffno+1.87*logFTNoninsstaffexp-0.29*logManagementexp-0.90*logCompEngScino+0.12*logHealthcareno-0.18*logServiceexp-0.82*logTotalFTEstaff+1.04*logCompEngSciFTE-1.75*logAllServiceFTE+0.71*logServiceFTE+1.01*logOfficeAdminFTE$
\end{itemize}
All variables are the same as the academic per capita model, and no
library variables enter the model.

\subsubsection{Combined binary per capita model}

The model selected in this case is:
\begin{itemize}
\item $logit(RDBINpc)=-117.51+1.08*logilltot+1.62*logRESSPACE-9.63*logCompEngScino+8.60*logCompEngSciexp+2.44*logCompEngSciFTE+1.20*logteachingotherinstrsupport-2.74*logLibcurarchteachingotherinstrsupportFTE$
\end{itemize}
Compared to the academic binary per capita model, \texttt{logNatResourceConstMaintFTE}
has been dropped and \texttt{logilltot} has entered the model. The
entry of \texttt{logilltot} into the model has reduced the coefficient
on \texttt{logRESSPACE}, while other coefficients have not changed
much. At least in this model, interlibrary loans and research space
are metrics that share some of the explanation for research funding.
Deviance residuals show no lack of fit, and overall deviance reduction
is moderate in this model.

\subsubsection{Combined clm per capita model}

Aside from slight changes in coefficients, the selected model in this
case is identical to the academic clm per capita model.
\begin{itemize}
\item $logit(RDCATpc)\approx-3.30*logTotalFTEstaff+2.05*logRESSPACE-8.52*logFTNoninsstaffno+8.01*logFTNoninsstaffexp+4.84*logCompEngSciFTE-4.43*logCompEngscino+0.92*logServiceFTE+1.95*logPHDResearch$
\end{itemize}
Looking at these last three combined models, we can see that the library
variables have little explanatory power when considering per capita
research output.

\section{Discussion}

\subsection{Comparison of models}

We summarize the variables selected by our models in a simplified
form to isolate the positive and negative effects of explanatory variables.
First, the models using ARL library data only: 

\medskip{}

\begin{tabular}{|c|c|}
\hline 
 & \emph{Library nominal}\tabularnewline
\hline 
\hline 
lm & logilltot - logsalstud + logtotexp + logphdawd + logphdfld + \tabularnewline
 & loggradstu - logexplm + logexpongoing - logtotstu\tabularnewline
\hline 
binary & loggradstu - logstudast + logexponetime + logprfstf\tabularnewline
\hline 
clm & logilltot + logphdfld + loggradstu - logstudast + logtotstfx\tabularnewline
\hline 
 & \tabularnewline
\hline 
 & \emph{Library per capita}\tabularnewline
\hline 
lm & logilltot + logsalprf + lognprfstf + logphdawd - logtotstfx\tabularnewline
\hline 
binary & logilltot + logsalprf - logtotstfx\tabularnewline
\hline 
clm & logilltot + logsalprf + logphdawd - logsalstud\tabularnewline
\hline 
\end{tabular}

\medskip{}

We see that the interlibrary loan variable, \texttt{logilltot}, enters
into all but one of the library models as a positive factor. Some
measure of the size of graduate programs, whether PhD's awarded or
graduate students, is nearly always present as a positive factor.
All per capita models show the salaries of library professionals as
a positive factor, whereas the nominal models tend to incorporate
variables for the overall size of staff and some variants of library
expenditure. Salaries of student workers and number of student workers
enter into the models as a negative factor for research funding, most
consistently in the per capita models.

In terms of complexity, the linear models include the most factors,
often with positive and negative factors in the same general area
(in the library case, positive effects for total expenditure and ongoing
expenditure and negative effects for materials expenditure). Per capita
models are more parsimonious than nominal models, with five variables
entering the per capita lm model. The four category clm model is simpler
with four variables, while the binary logistic model has the fewest
explanatory variables at three. 

Second, the models developed using academic indicators from NCSES
and IPEDS are presented below:

\medskip{}

\begin{tabular}{|c|c|}
\hline 
 & \emph{Academic nominal}\tabularnewline
\hline 
\hline 
lm & logRESSPACE + logCompEngSciFTE + logEndowment + logPHDResearch + \tabularnewline
 & logCompEngSciExp +logBusFinOpsFTE - logNatResourceConstMaintFTE - \tabularnewline
 & logCompEngScino - logMgmtFTE - logBusFinOpsexp \tabularnewline
\hline 
binary & logRESSPACE+ logTotalFTEstaff + logEndowment - logOfficeAdminFTE -
logMgmtFTE \tabularnewline
\hline 
clm & logRESSPACE +logTotalFTEstaff + logEndowment\tabularnewline
 &  - logAllServiceinclsalesofficeadminconstrmaintprodtransFTE - logMgmtFTE\tabularnewline
\hline 
 & \tabularnewline
\hline 
 & \emph{Academic per capita}\tabularnewline
\hline 
lm & logRESSPACE + logPHDResearch + logFTNoninsstaffexp + logHealthcareno
+ \tabularnewline
 & logCompEngSciFTE +logServiceFTE + logOfficeAdminFTE - logManagementexp
- \tabularnewline
 & logCompEngScino -logServiceexp -logTotalFTEstaff - logFTNoninsstaffno
- \tabularnewline
 & logAllServiceinclsalesofficeadminconstrmaintprodtransFTE \tabularnewline
\hline 
binary & logRESSPACE + logCompEngSciexp + logteachingotherinstrsupportFTE + \tabularnewline
 & logCompEngSciFTE - logCompEngScino - logNatResourceConstMaintFTE -\tabularnewline
 & logLibcurarchteachingotherinstrsupportFTE\tabularnewline
\hline 
clm & logRESSPACE + logFTNoninsstaffexp + logCompEngSciFTE + logServiceFTE
+\tabularnewline
 &  logPHDResearch - logTotalFTEstaff - logFTNoninsstaffno - logCompEngScino \tabularnewline
\hline 
\end{tabular}

\medskip{}

The models here explain much more of the variation in research, but
are also more complex. The linear models appear to overfit the data.
There are numerous parallel positive and negative terms for the same
employment categories, with number, expense, and FTEs receiving different
signs. However, research space, endowment, and research PhD's granted
are consistently positively associated with research funding. Staffing
relationships are complex, but we can note that computing, engineering,
and science staffing plays a large role in the per capita models.
Various other categories of employment such as management, sales,
and service, make their appearance as negative correlates of grant
funding. The overall picture supports the view that research intensity
is associated with research-centric inputs. 

As with the library variables, the clm and binary models produce simpler
equations with fewer significant variables. These models are easier
to interpret. For example, the nominal binary logistic model predicts
that endowment, research space, and total FTE staff (less office administrative
and management staff) are positively associated with research funding.
This is an intuitive and simple relationship.

Finally, the models developed using the combined set of variables
are summarized below:

\medskip{}

\begin{tabular}{|c|c|}
\hline 
 & \emph{Combined nominal}\tabularnewline
\hline 
\hline 
lm & logRESSPACE + logEndowment + logCompEngSciFTE + logPHDResearch + \tabularnewline
 & logCompEngSciExp +logBusFinOpsFTE - logNatResourceConstMaintFTE - \tabularnewline
 & logCompEngScino - logMgmtFTM - logBusFinOpsexp \tabularnewline
\hline 
binary & logRESSPACE + logEndowment+loggradstu - logstudast + logexponetime
- logMgmtFTE \tabularnewline
\hline 
clm & logRESSPACE + logEndowment+loggradstu\tabularnewline
\hline 
 & \tabularnewline
\hline 
 & \emph{Combined per capita}\tabularnewline
\hline 
lm & identical to Academic per capita - no library variables\tabularnewline
\hline 
binary & similar to Academic per capita, but with + illtot instead of logNatResourceConstMaintFTE\tabularnewline
\hline 
clm & identical to Academic per capita - no library variables\tabularnewline
\hline 
\end{tabular}

\medskip{}

The combined models show little effect of library variables. Some
of the ARL measures of institutional size enter into the models, but
the only variables about library activity that enter are one-time
expenses (in the nominal binary mode) and interlibrary loan (in the
per capita binary). Otherwise, the models are very similar to the
models selected from the academic-only variables.

\subsection{Findings}

Here we discuss the conclusions that can be drawn from the analysis
above, keeping in mind the caveat that this study is only a snapshot
of a single point in time for a limited number of institutions based
on available data, and that correlation does not imply causality.
\begin{itemize}
\item The library models provide some evidence that professional librarian
staffing is correlated with high levels of research activity. The
most consistent effect among library-specific variables is the positive
relationship of interlibrary loan levels to research output. Material
is not lent out via ILL unless it is in demand by the research community
and unique to the holding library. Duy and Lariviere \citep{DuyLariviere}
have studied the connection between ILL and research in the Canadian
context. Also, Henderson \citep{Henderson2000} has proposed a collection
failure quotient that takes interlibrary borrowing requests as a main
indicator of collection failure. These articles both argue for the
centrality of ILL as a measure of the distinctive strengths of an
institution's collection as opposed to the more crude title and volume
counts. Having high ILL rates is then an influential marker of the
quality of the library's collection, and its ability to support research
activity. ILL is the only library-specific variable to enter into
any of the per capita combined models. The variable for interlibrary
borrowing (\texttt{ilbtot}) might reflect faculty needs beyond a library's
holdings. However, interlibrary borrowing is not selected for in any
of the models, either as a positive of negative factor relating to
research. 
\item On the other hand, the fact that other library variables drop out
of the combined models means that larger claims about the library\textquoteright s
value to researchers are not directly verified by this study. By eliminating
effects purely related to institutional size, the per capita combined
models provide the best overall picture of the main linkages to grant
funding. In that case, high levels of research funding are associated
with the inputs most closely connected to the research itself: space,
staffing, and doctoral students. In two of the six combined models
(nominal and per capita binary models), the number of library variables
included were limited, and did not comprise what would normally be
considered primary measures of activity such as expenses, collection
size, and staffing. In the remaining four of the six combined models
(the nominal and per capita linear and cumulative logit models), measures
of library activity do not have any explanatory relationship to research
funding in per capita models. 
\item The fact that the library-variable only models explain much less of
the variation in research funding than the academic-variable models
(and the combined models) also argues for a weak relationship between
library strength, at least as currently measured, and research output.
\item The amount of research space available is important across all academic
and combined models. Since this is the most direct input into future
research, this effect is not surprising. To the extent that research
space is endogenous to grant funding success (with labs and facilities
constructed by previous grants), its importance as a predictor must
be tempered.
\item Endowment and other size-related measures are important predictors
in the nominal models, but staffing variables, especially in computing,
science, and engineering (STEM support), become more significant in
the per capita models.
\item The number of research PhD's granted is a significant positive factor
in most models, demonstrating research intensity more effectively
than numbers of faculty, master's students, or other measures of academic
activity.
\item Among regression methods, the linear models are accurate, but may
overfit the data. By including too many predictors, the nature of
the effects of each predictor are less easily understood. The categorical
approaches simplify prediction and understanding of effects, and avoid
overfitting. The clm models for multi-category data are midway in
complexity between the binary models and the linear regression models,
providing more meaningful and granular options for the response variable
while still yielding parsimonious variable selection. The final choice
among these models would depend on the desired level of granularity
in the response variable, research, versus the desire for a simpler
explanatory model. For many situations, the clm models may provide
the best balance among these requirements.
\item The variables that are not selected for in any models are also notable.
Traditional measures of library strength such as the number of volumes
held or number of unique titles do not appear in the models. Newer
measures such as e-books held or search counts do not appear, and
neither does the assistance offered by the library, as measured by
instructional sessions or reference questions answered. While one
may argue that these factors are related to student learning and success,
this study does not demonstrate that they are primary explanatory
factors for success in obtaining research funding.
\end{itemize}

\subsection{Conclusion and Extensions}

This study has gained some insight into the correlates of research
funding by examining one measure of research funding and its explanatory
variables in the limited population of US ARL institutions at one
point in time, 2012. By focusing directly on research output and considering
a wide range of variables and modeling approaches, this study provides
broader understanding of the relation of various factors to high levels
of grant funding. Weighing library variables alongside non-library
variables is a sounder way of assessing library impact than looking
at library measures in isolation. However, in this case, we find only
a few significant library relationships to research. Logistic regression
models have not been used in prior library literature to investigate
such issues, and this study shows that categorical data representations
of continuous variables, used with logistic and cumulative logit modeling,
have advantages in producing simpler, easier to understand models
with more significant main effects. 

The limited population and single time period of study limit the generalizability
of the findings presented here, but there are many potential extensions
of this research.

One direction of expansion would be to study trends over time by looking
at longitudinal data for changing causal relationships. Another direction
would be to look at a wider selection of libraries. As mentioned in
the introduction, the ARL institutions, while representative, are
not a complete set of the major research institutions. IPEDS has data
on all academic libraries. While it is not as frequently collected
or quite as comprehensive, it could be used for library metrics from
a much larger group of libraries. This data could also be used to
compare research funding correlates at smaller institutions. 

The most promising immediate extension of this research would be to
use the full detail present in the NCSES HERD survey, which contains
breakdowns of federal funding, funding by agency (such as NSF), and
funding by subject discipline. NSF funding measures would represent
a broad base of general scientific research, but would also avoid
some of the data issues mentioned in the introduction concerning the
sometimes separate and sometimes merged medical research and library
units in the modern university. Studying specific disciplines would
also reveal the unique characteristics of each.

Also, the per capita analysis in this project converted only the response
variable to a per capita basis. At the cost of generating many other
variables to consider, one could convert many of the explanatory variables
to a per capita basis, such as library expenditures per faculty or
research space per faculty. These measures may generate models with
different implications. This is certainly worth pursuing to provide
a more thorough analysis.

Other methodological refinements to the regression models presented
here may produce more robust results.

Those are all possible future directions for research. This study
has taken an initial step in demonstrating that linear, logistic,
and cumulative logit models, when combined with a broad selection
of data representing many aspects of the academic enterprise can be
used to explain the correlates of research funding at US ARL institutions.

\pagebreak{}

\bibliographystyle{plainurl}
\bibliography{/home/ryan/womack/documents/ryan/BibTeX/Stat553}

\pagebreak{}

\section*{Appendix on Data Cleaning}

While the data used is mostly presented as it appears in the original
data sources, there are some adjustments made for consistency. 

For some universities, NSF data for Health Sciences and Medical units
are reported separately. If these units were in the same geographic
location as the main campus, their data was added to the main campus
totals. However, 2012 is prior to the integration of Rutgers University
with the University of Medicine and Dentistry of New Jersey (UMDNJ),
so the UMDNJ data is not added to Rutgers. 

In order to remove the possibility of having negative infinity as
the result of a log transform, occurrences of \textquotedblleft 0\textquotedblright{}
in the data were modified to \textquotedblleft 1\textquotedblright .
The magnitudes of the variables were much higher across the board,
with results in hundreds or thousands, so \textquotedblleft 1\textquotedblright{}
can be viewed as \textquotedblleft almost zero\textquotedblright{}
in this context. 

The University of Colorado does not report endowment separately by
campus, so the endowment of the University of Colorado System including
all branches is substituted. Since there are no other campuses in
the system that rival the history and research success of the Boulder
campus, this is not likely to introduce much distortion. 

The IPEDS data classifies teachers into \textquotedblleft instructional\textquotedblright{}
and \textquotedblleft research\textquotedblright{} teaching staff,
but the reported results are very inconsistent, with some institutions
having no instructional and others having no research staff, so these
variables were not used in the analysis. 

In the ARL data, some data recorded as \textquotedblleft 0\textquotedblright{}
actually appears to be missing, since it is unreasonable to think
that at a large university, there are no presentations, reference
transactions, PhD\textquoteright s awarded, and so on. The entries
that have been converted to missing are summarized in Table 7. 

\begin{table}[t]
\protect\caption{Library ``Zero'' data converted to Missing}

\begin{tabular}{|c|c|}
\hline 
variable & Institutions\tabularnewline
\hline 
\hline 
group presentation & Washington State\tabularnewline
\hline 
group pres. participants & Washington State\tabularnewline
\hline 
reference transactions & Rice, Maryland, Pennsylvania, Wisconsin\tabularnewline
\hline 
student assistants & Harvard{*}\tabularnewline
\hline 
Ph.D's awarded & Washington State\tabularnewline
\hline 
titles (\# in library) & Pittsburgh\tabularnewline
\hline 
one-time expenses & Cornell, Georgetown\tabularnewline
\hline 
ongoing expenses & Cornell, Georgetown\tabularnewline
\hline 
collection support expenses & Cornell, Georgetown, Georgia Tech, \tabularnewline
 & UC-San Diego, UC-Santa Barbara\tabularnewline
\hline 
 & \emph{{*}Harvard Library website does note student employment}\tabularnewline
\hline 
\end{tabular}
\end{table}

The variable for full text article requests has 11 institutions reporting
zeroes. This is too many missing values for our small data set, so
we omit this variable from the analysis. 

In general, status variables (e.g., presence of a Hospital) are coded
\textquotedblleft 1\textquotedblright{} for yes, \textquotedblleft 0\textquotedblright{}
for no. In addition, institutional control is coded \textquotedblleft 0\textquotedblright{}
for private, \textquotedblleft 1\textquotedblright{} for public. 
\end{document}